\documentstyle[12pt]{article}
\begin{document}
\setlength{\parskip}{0.45cm}
\setlength{\baselineskip}{0.75cm}
%
%
%
\begin{titlepage}
\setlength{\parskip}{0.25cm}
\setlength{\baselineskip}{0.25cm}
\begin{flushright}
DO-TH 99/17\\  
\vspace{0.2cm}
\end{flushright}
\vspace{1.0cm}
\begin{center}
\LARGE
{\bf Rescattering Effects for $(\varepsilon\,'/\varepsilon)$}
\vspace{1.5cm}

{\large
Emmanuel A.\ Paschos}\\

\vspace{1.0cm}
\normalsize
{\it Institut f\"{u}r Physik, Universit\"{a}t Dortmund}\\ 
{\it D-44221 Dortmund, Germany} \\

\vspace{1.5cm}
\end{center}
\begin{abstract}
\noindent A calculation of the hadronic matrix elements for $K\to\pi\pi$ 
including 
$\pi$--$\pi$ \mbox{rescattering} effects in a
dispersion integral is presented.  I study the dependence of the 
results on the
matching scale $\mu$ and use them to calculate the CP--parameter
$(\varepsilon\prime/\varepsilon)$.
I find improved stability on the matching scale and good agreement
with the experimental results.
\end{abstract}
\end{titlepage}
%
During the past few years there have been new developments on the 
problem of direct CP--violation.  On the theoretical side there are
predictions \cite{ref1}--\cite{ref5} which agree with the new experimental
results \cite{ref6,ref7} and some calculations which are still far
away from the data.  The agreement between theory and experiments must
be considered a success of QCD since the dominant contributions to  
$(\varepsilon\prime/\varepsilon)$ come from the effective Hamiltonian
generated through renormalization of the weak interaction.  The
renormalization coefficients were computed by two groups \cite{ref8,ref9}
and agree with each other. 

It is clear now that the original calculations of the matrix elements,
done in the Vacuum Saturation Approximation (VSA), must be supplemented
by rescattering corrections of the two pions, which increase 
$\langle Q_6\rangle_0$ and decrease $\langle Q_8\rangle_2$.
The rescattering corrections were described in an earlier calculation
\cite{ref1,ref2} which made the ratio  
$(\varepsilon\prime/\varepsilon)$ positive.  A new and improved 
calculation in the
chiral theory of pseudoscalar mesons \cite{ref3,ref4} and the chiral
quark model \cite{ref5} obtained results consistent
with large values for $(\varepsilon\prime/\varepsilon)$.  In this article
I introduce a new method for calculating the rescattering corrections,
which may also be useful for other low--energy calculations.  The 
results indicate that $\pi$--$\pi$ rescattering brings important
corrections and allow to compute the imaginary parts of the 
amplitudes.

The operators occurring in the effective Hamiltonian are local
operators which satisfy analyticity and threshold conditions of field
theory.  Consequently their matrix elements can be written as integrals
over their singularities.  There are two types of contributions to 
each matrix element:  tree diagrams and loops.  The four dimensional
integrals for the loops can be rewritten as one--dimensional integrals
over the energy flowing through the K--meson.  We split the integrals
in two regions -- from $4m_{\pi}^2$ to $\mu^2$ and from $\mu^2$
to infinity.  The higher
energy region is computed in QCD and its discontinuity is related to
the anomalous dimension.  The low energy part can be represented by an
effective theory or by experimental data.  The final formula for the
amplitude at a low energy point $\sigma=m_K^2$ can be written as
\newpage
\begin{eqnarray}
Re\, A(\sigma) & = & {\Large\sl a} + \frac{1}{\pi} \int_{4\pi^2}^{\infty} 
    \frac{Im\, A(s)}{s-\sigma}\, ds = {\Large\sl a} 
       + \frac{1}{\pi} \int_{4m_{\pi}^2}^{\mu^2}\frac
	{Im\, A(s)}{s-\sigma}\, ds +{\Large\sl a}\gamma\,
         \frac{\alpha_s}{4\pi}\,ln 
          \frac{M_t^2}{\mu^2} + \ldots\nonumber\\[3mm] 
     & = &  C(\mu) \left \{ {\Large\sl a} +\frac{1}{\pi} 
              \int_{4m_\pi^2}^{\mu^2}
               \frac{Im\, A(t)}{s-\sigma}\, ds\right \}.
\end{eqnarray}

\noindent Here ${\Large{\sl a}}$ is the lowest order contribution, which 
can be 
substituted by
the vacuum saturation or another improved approximation.  In the last
equation we factorized two terms and identified one of them with
the series generated by QCD.  We emphasize that we do not need a 
subtracted dispersion integral because for the high--energy region
we use directly the QCD realization of the theory.  This is a useful
expression for the low--energy value of the amplitude and may have
applications beyond the cases described in this article.  In addition,
eq.\ (1) provides a matching between the two regimes at the cutoff
scale $\mu$, which can be varied from 0.8 GeV all the way up to 1.8 GeV
where the validity of QCD is more reliable.
 
In the $K\to 2\pi$ decays we have many matrix
elements of the operators, generated through QCD, but very few observables.
To improve the situation we
shall try to calculate the amplitudes and their strong phases.  For 
the $A_2$ amplitude we use the value
\begin{equation} A(K^+\to \pi^+\pi^0) = (1.837 \pm 0.002)\cdot 10^{-8}\,\,
  {\rm{GeV}}
\end{equation}
and its phase is given, according to Watson's theorem, by the phase 
extracted from $\pi$--$\pi$ scattering.  In fact we will calculate the
phase for each matrix element and compare it with the $\pi$--$\pi$ phase
shift.  
The imaginary part for each rescattering is given through unitarity
by the cut diagram shown in figure 1; where the square is a weak vertex
and the circle a strong vertex.  We obtain

\unitlength1cm
\begin{figure}[ht]
\begin{center}
\begin{picture}(8.6,0.5)
\put(0,0.25){\line(1,0){2.5}}
\thicklines \put(2.5,0.0){\framebox(0.5,0.5)[t]{}} \thinlines
\put(3.75,0.25){\oval(1.5,1.5)[l]}
\put(4.75,0.25){\oval(1.5,1.5)[r]}
\thicklines \put(5.75,0.25){\circle{0.5}} \thinlines
\put(6.75,0.25){\oval(1.5,1.5)[l]}
\put(0.5,0.5){$K$}
\put(3.5,1.3){$\pi$}
\put(3.5,-1.0){$\pi$}
\put(4.7,1.3){$\pi$}
\put(4.7,-1.0){$\pi$}
\put(6.6,1.3){$\pi$}
\put(6.6,-1.0){$\pi$}

\end{picture}
\vspace{1cm}
\caption{}
\end{center}
\end{figure}

\vspace{-1.0cm}
\begin{equation}
Im\langle\pi\pi,\, I=0|Q_6|K^0\rangle = \langle\pi\pi,\, I=0|Q_6|K^0
  \rangle_{VSA}\, \frac{1}{4}\, \frac{m_K^2-\frac{1}{2}m_{\pi}^2}{4\pi\,
    F_{\pi}^2}\, \left( 1 - \frac{4m_{\pi}^2}{m_K^2}\right)^{1/2}
\end{equation}
and
\begin{equation}
Im\langle\pi\pi,\, I=2|Q_8|K^0\rangle = \langle\pi\pi,\, I=2|Q_8|K^0
  \rangle_{VSA}\, \left(-\frac{1}{8}\right)\, \left(\frac{m_K^2-2m_{\pi}^2}
   {4\pi\,F_{\pi}^2}\right)\, \left( 1-\frac{4m_{\pi}^2}{m_K^2}\right)^{1/2}
\end{equation}
I shall use these equations for computing the imaginary parts throughout
this paper.  Using them we computed the numerical values for
$Im\langle Q_8\rangle_2$ in Table (1) of reference \cite{ref4}.  The 
imaginary part of $\langle Q_6\rangle_0$ was not included in \cite{ref4}
because it is higher order in the expansion described in that article. 
The form of the equations
indicates that the imaginary part consists of two multiplicative factors:
a weak vertex and the $\pi$-$\pi$ scattering amplitude characterized,
in our case, by two isospin states $I=0,\, 2$.

We proceed now to calculate the real part of the amplitudes.  For this
calculation we need the imaginary part for values of the center--of--mass
energy squared, s, in the range $4m_{\pi}^2\leq s<\mu^2$.  The proof
of Watson's theorem holds for the matrix element of each operator
and their imaginary parts are given as
\begin{equation}
Im\, A_I = Im\langle\pi\pi,\, I|Q_i|K^0\rangle = |\langle\pi\pi,\, I|Q_i|
    K^0\rangle| \cdot \sin\delta_{\ell=0}^I(s)
\end{equation}
with $\delta_0^I$ the experimental phase shifts for isospin $I=0,\,2$
pion--pion scattering.  
The imaginary part is given by this formula in the elastic region and
I adopt this form beyond the elastic region using the experimental
phase shifts.  For the magnitude of the matrix element we
can take the low energy contribution, mentioned earlier, but we are free to
introduce a weak energy dependence; for instance the variation introduced
by the real part obtained through the dispersion relation.
Let us denote by
\begin{equation}
A_{Q_{i,I}}(s) = \langle \pi\pi,\, I|Q_i|K^0\rangle.
\end{equation}
Then
\begin{equation}
Re\, A_{Q_{i,I}}(\sigma) = \left\{ {\Large\sl a}_{Q_i} +\frac{1}{\pi}
   \int_{4m_{\pi}^2}^{\mu^2}|
    \langle \pi\pi,\, I|Q_i|K^0\rangle|\, \frac{\sin\delta_0^I(s)}
     {s-\sigma}\, ds\right\} \, C_i(\mu).
\end{equation}
We shall assume that the absolute value of the matrix element is a slowly
varying function of energy over the region of integration and we use 
the experimental values for $sin\delta_0^I$ to perform the principal
value integral.  We define the functions
\begin{equation}
f_I(\mu,\sigma) = \frac{1}{\pi}\int_{4m_{\pi}^2}^{\mu^2}\, \frac
  {\sin\delta_0^I(s)}{s-\sigma}\, ds
\end{equation}
and calculated the values presented in table 1.  These terms bring in a 
correction to the tree level contribution.  For the integrations we use
the experimental phase shifts from references \cite{ref10}--\cite{ref13}.
Data for $\delta_0^2$ exist up to the energy of 1.5 GeV and we performed
the integral up to this value.  For $\delta_0^0$ the data show clearly
the $\rho$--resonance.  They extend to 
1.8 GeV and I give the additional values in the table.  The functional
change of the phase shift given in \cite{ref11} produces an $f_0(\mu,m_k)$
which is almost constant for $\mu>900$ MeV.  The extended ranges presented
in table 2 allow us to study the dependence of $Re A_{Q_{i,I}}$ on the 
matching scale $\mu$.  As an additional test, I introduced a 20\% energy
dependence on the magnitude of the matrix element and carried out the
integration.  The results differ only by one or two units in the second
decimal.

As a first test of the approach we consider the $A_2$ amplitude.
Its Born contribution is known
\begin{equation}
A_2^{\rm{Born}}(K^+\to\pi^+\pi^0) = \sqrt{\frac{3}{2}}\, \frac{G_F}
     {\sqrt{2}}\, V_{ud}\, V_{us}^*(z_1+z_2)\langle Q_i\rangle_2 
       = 2.74 \times 10^{-8}\,\,{\rm{GeV}}.
\end{equation}
\begin{center}
\begin{tabular}{|c|c|c|}
\hline
$\mu$ in GeV & $f_2(\mu,m_k)$ & $f_0(\mu,m_k)$\\[1mm]
\hline\hline
0.7 & -0.09 & 0.34\\[1mm]
0.8 & -0.12 & 0.50\\[1mm]
0.9 & -0.17 & 0.60\\[1mm]
1.1 & -0.23 & 0.71\\[1mm]
1.3 & -0.29 & 0.64\\[1mm]
1.5 & -0.34 & 0.55\\[1mm]
1.7 & ---   & 0.55\\[1mm]
1.9 & ---   & 0.60\\[1mm]
\hline
\end{tabular}\\

\vspace{0.5cm}
{Table 1: Numerical results for the principal value integrals}
\end{center}
\vspace{-0.2cm}
Including next the unitarity corrections, at $\mu=0.9$ GeV we obtain
\newpage
\begin{eqnarray}
A_2^{\rm{complete}}(K^+\to\pi^+\pi^0) & = & 2.74 (1 - 0.17-i 0.20)
            \cdot 10^{-8}\nonumber\\
& = & (2.27 - i 0.55)\cdot 10^{-8}\,\,{\rm{GeV}}\\
\tan\theta & = & -0.24,\quad\quad\quad \theta = -13.7^o.
\end{eqnarray}
We note that the magnitude of the calculated amplitude is larger than the
measured value by 30\%.  The phase of the amplitude has the correct sign
and it is slightly larger than the experimental value.  The experimental
values reported in the articles \cite{ref10}--\cite{ref13} vary among
themselves.  An approximate value is $-10.5\pm1.6^0$.  A more accurate
value was obtained through a dispersion calculation for $\pi$--$\pi$
scattering \cite{ref11}, but the error quoted is very small and is perhaps
an underestimate.  Finally, the dependence of the amplitude on the 
matching scale $\mu$ is small, as discussed for the other two matrix
elements below.  

The calculation for $A_{Q_{8,2}}$ proceeds along similar lines.  We
obtain the imaginary part from eq.\ (4) and the real part from eq.\ (7)
and table 1
\begin{equation}
A_{Q_{8,2}}(m_k) = \langle\pi\pi,\, I=2|Q_8|K^0\rangle_{\rm{VSA}}
  (1 - 0.20 - i 0.20)\,y_8(\mu)
\end{equation}
at $\mu=1.0$ GeV and with \cite{ref3,ref4}
\begin{equation}
\langle\pi\pi,\, I=2|Q_8|K^0\rangle_{\rm{VSA}} = \frac{\sqrt{3}}
  {2\sqrt{2}}\, r^2\, F_{\pi}\, \left[ 1+ \frac{8m_k^2}{F_{\pi}^2}\,
    (L_5-2L_8) - \frac{4m_{\pi}^2}{F_{\pi}^2}\, (3L_5-8L_8)\right]
\end{equation}
and $r=\frac{2m_k^2}{m_s+\hat{m}}$.  The renormalized couplings $L_5$
and $L_8$ are defined in references \cite{ref3,ref4} and have the
numerical values
\begin{equation}
L_5=2.07\cdot 10^{-3}\quad\quad{\rm{and}}\quad\quad L_8=1.09\cdot 10^{-3}.
\end{equation}
Substituting the numerical values for $m_s\, (1\, {\rm{GeV}})=150$ MeV 
we obtain
\begin{equation}
A_{Q_{8,2}}(m_k) = (0.37 - i 0.09)\, y_8{(\mu)}\,\,{\rm{GeV}}^3.
\end{equation}
The phases in eqs.\ (10) and (15) are the same, which is an attractive
property of the dispersion relation, i.e.\ all $I=2$ matrix elements have
the same phase.  A similar property holds for the $I=0$ matrix elements.
Finally, we can investigate the dependence on the matching scale $\mu$
which appears on the factor $r^2$, through the running mass of the strange
quark, and the Wilson coefficient $y_8(\mu)$. In a specific regularization
scheme the product $r^2y_8(\mu)$ is stable between 1 and 2 GeV, varying
by less than 5\%.  The variation among the three regularization schemes
in the same energy region \cite{ref14} is at most 15\%.

The unitarity corrections to the matrix element $Q_6$ are controlled by
the phase shift $\delta_0^0$ which is positive, giving a positive 
correction for the real part of the matrix element.  The increase of the
matrix element is given by the function $f_0(\mu)$ in table 1, which is
stable above 900 MeV.  At $\mu=1$ GeV
\begin{eqnarray}
A_{Q_{6,0}} & = & \langle \pi\pi|Q_6|K^0\rangle_{\rm{VSA}}\, 
  (1 + 0.60 + i 0.40)\, y_6(\mu)\nonumber\\
& = & (-0.56 - i 0.14)\, y_6\, (\mu)\,\,{\rm{GeV}}^3
\end{eqnarray}
with \cite{ref3,ref4}
\begin{equation}
\langle \pi\pi|Q_6|K^0\rangle_{\rm{VSA}} = -4\sqrt{3}\, r^2\L_5\,
  \frac{m_k^2-m_{\pi}^2}{F_{\pi}}.
\end{equation}
Again the product $r^2y_6(\mu)$ is very stable for $1<\mu<2$ GeV;
the variations mentioned in the previous paragraph for $r^2y_8(\mu)$
again hold.  The phase of this amplitude is positive and equal to
$+14^o$ which is approximately half the experiment phase shift at
$\sqrt{s} = m_k$. 

With the values derived already we can compute the parameter
$(\varepsilon\prime/\varepsilon)$.  The standard derivation leads
to the expression
\begin{equation}
\frac{\varepsilon\prime}{\varepsilon}
 = \frac{G_F}{2}\, \frac{\omega}{|\varepsilon|ReA_0}\, Im\,\lambda_t
  \left[ \pi_0 -\frac{1}{\omega}\pi_2\right]
\end{equation}
\vspace{-0.7cm}

with
\vspace{-1.5cm}
\begin{eqnarray}
\pi_0 & = &|\Sigma_i\, y_i(\mu)\langle Q_i\rangle_0|(1-
   \Omega_{\eta\eta\prime})\\
\pi_2 & = &|\Sigma_i\, y_i(\mu)\langle Q_i\rangle_2|\\
Y     & = & \pi_0 -\frac{1}{\omega}\, \pi_2
\end{eqnarray}
and $\Omega_{\eta\eta\prime}\sim 0.25\pm 0.05$ being the isospin breaking in
the quark masses ($m_u \neq m_d$).  The absolute values originate from
the fact that the phases of strong origin were already extracted in the
calculation of $\varepsilon\prime/\varepsilon$.  The overall factor is
precisely known $\frac{G}{2}\, \frac{\omega}{|\varepsilon|ReA_0} =
346$ GeV$^{-3}$ and the Cabbibo--Kobayashi--Maskawa factor was recently
estimated \cite{ref14}.
\begin{equation}
Im\, \lambda_t = (1.38\pm 0.33)\times 10^{-4}
\end{equation}
This new value is a large improvement over the values reported in the
early 90's and leads to a large reduction of the uncertainties.
A second reduction of uncertainties comes from the weak dependence of 
the amplitudes in
eqs.\ (12), (15) and (16) on the matching scale.
We summarize in table 2 the uncertainties for $(\varepsilon\prime/
\varepsilon)$ originating from two sources.
\begin{center}
\begin{tabular}{|c|l|}
\hline
$\quad$ source $\quad$ & 
$\quad\Delta(\varepsilon\prime/\varepsilon)\quad$\\[1mm]
\hline\hline

$\quad Im\, \lambda_t\quad$ & $\quad\pm 25\%$\quad\\[1mm]
$\quad$ Matching $\quad$        & $\quad\pm$ 20\%\quad\\[1mm]
\hline
\end{tabular}\\

\vspace{0.5cm}
{Table 2: Theoretical Uncertainties}
\end{center}
\vspace{-0.2cm}
Another uncertainty comes from the strange quark mass which enters the
calculation of the matrix elements through the factor $r=\frac{2m_k^2}
{m_s+\hat{m}}$.  Values for the running strange quark mass have been
computed by various methods and vary considerably.  Older estimates
\cite{ref15} and QCD sum rules \cite{ref16} give higher values 
$m_s(1\, {\rm{GeV}})\approx 150\pm 55$ MeV at the usual renormalization
point $\mu=1$ GeV;
while recent values from lattice calculations \cite{ref17} give 
smaller values $\approx 110\pm20$ MeV at $\mu=2$ GeV.  
It is customary to adopt the range \cite{ref4}
\begin{equation}
m_s(1\, {\rm{GeV}}) = 150 \pm 25\, {\rm{MeV}}.
\end{equation}
For central values of the parameters at 1 GeV and keeping the
errors from table 2 in quadrature, I obtain
\begin{eqnarray}
\varepsilon\prime/\varepsilon & = & (4.2 \pm 1.3)\cdot (0.0320)
     \cdot 10^{-2}\nonumber\\
& = & (15.0\pm 4.8)\cdot 10^{-4},
\end{eqnarray}
where the number 0.0320 comes from $\langle Q_6\rangle_0$
including $\Omega_{\eta\eta\prime}$ minus the contribution from
$\langle Q_8\rangle_2$. I kept only these two operators and found out
that the contribution from the $I=2$ amplitude is only $20\%$ of the
$\langle Q_6\rangle(1-\Omega_{\eta\eta\prime})$ term. This value for 
the ratio is in good agreement with the average experimental value
$(\varepsilon\prime/\varepsilon) = (21.2\pm 4.6)\cdot 10^{-4}$.

I have shown in this article that the VSA for the matrix elements 
together with rescattering corrections lead to values of 
$(\varepsilon\prime/\varepsilon)$ which are consistent with the 
experimental measurements.  The unitarity corrections improve the
stability on the matching scale $\mu$.  The specific values of the
phase--shifts increase $\langle Q_6\rangle_0$ and decrease $\langle
Q_8\rangle_2$ making the difference in the function $Y$ positive
definite \cite{ref1} -- a feature which is maintained in many
calculations \cite{ref2}--\cite{ref5}, \cite{ref19}.  Two very recent
articles \cite{ref18,ref19} use dispersion relations for the K--meson
decay amplitudes. 
They differ in several basic respects from the 
present article and the interested reader can study them for 
comparison. 
Finally, it will be interesting to test if lattice calculations
\cite{ref20,ref21}, which fullfil unitarity corrections, still give 
different results.   

A more extensive
exposition of this work including numerical studies and other 
applications
will be presented in the future.  In particular, I wish to study the
various contributions to the 
$\Delta I=\frac{1}{2}$ rule and compare them with the results 
of previous calculations \cite{ref22}--\cite{ref24}.
\vspace{1.5cm}

\noindent{\Large{\bf{Acknowledgement}}}

\noindent This work has been supported in part by the 
`Deutsche Forschungsgemeinschaft' (DFG), Bonn (Pa-254/10-1), and  
by the `Bundesministerium f\"ur
Bildung, Wissenschaft, Forschung und Technologie', Bonn (05HT9PEB8).
I wish to thank Drs.\ T.\ Hambye, \mbox{G.\ K\"ohler}, P.\ Soldan and 
Y.-L.\ Wu for discussions and Dipl.--Phys.\ W.\ Rodejohann for 
discussions and help with the numerical work.


\vspace{1.5cm}


\begin{thebibliography}{13}
\bibitem{ref1}  J.\ Heinrich, E.A.\ Paschos, J.-M.\ Schwarz and Y.L.\ Wu,
		{\it Phys.\ Lett.} {\bf B279}, 140 (1992)  
\bibitem{ref2}  E.A.\ Paschos, Invited Talk at the 17th Lepton--Photon
		Symposium, Beijing, China (Aug.\ 1995)
\bibitem{ref3}	T.\ Hambye, G.O.\ K\"ohler, E.A.\ Paschos, P.\ Soldan and
		W.A.\ Bardeen, {\it Phys.\ Rev.} {\bf D58}, 014017 (1998)
\bibitem{ref4}	T.\ Hambye, G.O.\ K\"ohler, E.A.\ Paschos and P.\ Soldan,
		eprint hep--ph 9906434, to be published in
		{\it Nucl.\ Phys.\ B}	
\bibitem{ref5}	S.\ Bertolini, J.O\ Eeg, M.\ Fabbrichesi and E.J.\ Lashin, 
		{\it Nucl.\ Phys.} {\bf B514}, 93 (1998)
\bibitem{ref6}  G.D.\ Barr et al.\ (NA31), {\it Phys.\ Lett.}
		{\bf B317}, 233 (1993);\\
		V.\ Fanti et al.\ (NA48), {\it Phys.\ Lett.} {\bf B465},
		335 (1999)
\bibitem{ref7}  L.K.\ Gibbons et al.\ (E731), {\it Phys.\ Rev.\ Lett.}
		{\bf 70}, 1203 (1993);\\
~		A.\ Alavi--Harati et al.\ (KTeV), {\it Phys.\ Rev.\ Lett.}
		{\bf 83}, 22 (1999)  
\bibitem{ref8}  A.\ Buras, M.\ Jamin, M.E.\ Lauterbacher and P.H.\ Weisz,
		{\it Nucl.\ Phys.} {\bf B400}, 37 (1993)
\bibitem{ref9}  M.\ Ciuchini, E.\ Franco, G.\ Martinelli and L.\ Reina,
		{\it Phys.\ Lett.} {\bf B301}, 263 (1993)
\bibitem{ref10} O.O.\ Patarakin, V.N.\ Tikhonov, K.N.\ Mukhin, 
		{\it Nucl.\ Phys.} {\bf A598}, 335 (1996)		
\bibitem{ref11} E.\ Chell and M.G.\ Olsson, {\it Phys.\ Rev.} {\bf D48},
		4076 (1993)
\bibitem{ref12} W.\ Hoogland et al., {\it Nucl.\ Phys.}
 		{\bf B126}, 109 (1977) 
\bibitem{ref13}	P.\ Eastbrooks and A.\ Martin, {\it Nucl.\ Phys.}
		{\bf B79}, 301 (1974)
\bibitem{ref14}	G.\ Buchalla, A.\ Buras and M.\ Lautenbacher,
		{\it Rev.\ Mod.\ Phys.} {\bf 68}, 1165 (1996)
\bibitem{ref15}	J.\ Gasser and H.\ Leutwyler, {\it Phys.\ Rep.} 
		{\bf 87}, 77 (1982);\\
		G.\ Dillon and G.\ Morpurgo, {\it Phys.\ Rev.}
		{\bf D53}, 3754 (1996)
\bibitem{ref16}	S.\ Bosch, A.J. Buras, M.\ Gorbahn, S.\ Jager, M.\ Jamin,
		M.E.\ Lautenbacher and \mbox{L.\ Silvestrini}, 
		eprint hep--ph/9904408
\bibitem{ref17}	R.\ Gupta, eprint hep--ph/9801412,\\
		J.\ Garden, J.\ Heitger, R.\ Sommer and H.\ Wittig,
		eprint hep--lat 9906013
\bibitem{ref18} E.\ Pallante and H.\ Pich, eprint hep--ph/9911233
\bibitem{ref19} J.F.\ Donoghue and E. Golowich, eprint hep--ph/9911309
\bibitem{ref20}	M.\ Ciuchini, E.\ Franco, G.\ Martinelli and L.\ Reina,
		{\it Phys.\ Lett.} {\bf B301}, 263 (1993);\\
		M.\ Ciuchini et al., hep--ph/9910237
\bibitem{ref21}	G.\ Kilcup, {\it Nucl.\ Phys.} {\bf B} (Proc. Suppl.)
		{\bf 73}, 417 (1991);\\
		T. Blum et al., BNL--preprint 66731 (1999), 
		hep--lat 9908025
\bibitem{ref22}	W.A.\ Bardeen, A.J.\ Buras and J.-M.\ Gerard,
		{\it Phys.\ Lett.} {\bf B180}, 133 (1986) and
		\mbox{{\it Phys.\ Lett.}} {\bf B192}, 138 (1987)
\bibitem{ref23}	T.\ Hambye, G.O.\ K\"ohler and P.H.\ Soldan,
		{\it Eur.\ Phys.\ J.} {\bf C10}, 271 (1999) 
\bibitem{ref24}	J.\ Bijnens, J.\ Prades, {\it Nucl.\ Phys.} {\bf B521},
		305 (1998)		
\end{thebibliography}
\end{document}